\documentclass[a4paper,8pt,oneside]{article}
\usepackage{subfigure}
\usepackage{graphicx}
\usepackage{cite}
\usepackage{multicol}
\usepackage{ulem}
\makeatletter

\newenvironment{fig}
{\def\@captype{figure}}
{}
\makeatother
\begin{document}
\title{First-principles calculation on the transport properties of molecular wires between Au clusters under equilibrium}
\author{Zhanyu Ning\footnote{Corresponding author. Email:zyning@pku.edu.cn}$^{\ ,1}$, Jingzhe Chen$^{1}$, Shimin Hou$^{2}$, Jiaxing Zhang$^2$, Zhenyu Liang$^{1}$,\\ Jin Zhang$^{1}$, and Rushan Han$^{1}$\\ \small{$^{1}$ School of Physics, Peking University, Beijing \ 100871, People's Republic of China}\\\small{$^2$ Department of Electronics, Peking University, Beijing\ 100871, People's Republic of China}} 

\date{}
\maketitle
\begin{abstract}
Based on the matrix Green's  function method combined with hybrid tight-binding / density functional theory, we calculate the conductances of a series of gold-dithiol molecule-gold junctions including  benzenedithiol (BDT), benzenedimethanethiol (BDMT), hexanedithiol (HDT), octanedithiol (ODT) and decanedithiol (DDT). An atomically-contacted extended molecule model is used in our calculation. As an important procedure, we determine the position of the Fermi level by the energy reference according to the results from ultraviolet photoelectron spectroscopy (UPS) experiments. After considering the experimental uncertainty in UPS measurement, the calculated results of molecular conductances near the Fermi level qualitatively agree with the experimental values measured by Tao et. al. [{\it Science} 301, 1221 (2003); {\it J. Am. Chem. Soc.} 125, 16164 (2003); {\it Nano. Lett.} 4, 267 (2004).] 
\end{abstract}

\section{Introduction}
Molecular devices are now playing an important role in nanoelectronic researches. The construction, measurement and understanding of the transport  properties in molecular electronic circuits have drawn more and more attention. At present, a lot of theoretical work is concentrating on the sandwichtype structure of an individual molecule between metallic electrodes \cite{NDLANG,xue1,xue2,damle}. Although current simulation methods can qualitatively describe some electronic properties of such simple systems, there are still a lot of issues waiting to be made understandable for the purpose of predicting, designing and controlling new molecular devices. For example, the methods used to calculate the coupling strength of electrodes-molecule interaction remain unconvincing. It's difficult to model the metal-molecule-metal junction accurately because of the uncontrollable interface structures between molecules and leads. In addition, the position of the Fermi energy is another focal point. The common simple viewpoint is that the Fermi level of electrodes lies within the gap of molecular orbitals between HOMO (highest occupied molecular orbital) and LUMO (lowest unoccupied molecular orbital) \cite{Ratner}. Further calculation and experimental results are essentially needed to clarify the doubt of precise location of the Fermi energy. On the other hand, the importance of the modification of molecules due to the non-equilibrium situations and the determination on the spatial profile of the electrochemical potential across the interface  also need more discussions. 

 For device under equilibrium, the above problems reduce to calculating electronic structures of the molecules sandwiched between two metal electrodes, which is the first step to understand the transport behaviors of molecular devices under any circumstances. Recently, the conductances of a series of gold-dithiol molecule-gold systems were measured near low bias by the tip of scan tunneling microscope (STM) \cite{tao1,tao2,tao3}. Though the molecular junctions based on Au-S bond have been investigated extensively \cite{xue3,xue4,cms27151,luoyi,prb65245105,jcp1218050}, our calculation and further discussions according to the experimental results still can show more enlightenments for understanding the interaction between molecules and metals. Ultraviolet photoelectron spectroscopy (UPS) was referenced for the purpose of determining ``line up'' about the Fermi level \cite{ups1,ups2,ups3,ups4,ups5}. UPS experiments can provide direct spectroscopic information about the gap between HOMO and $E_F$, which is one of the key points in the transport calculation of molecule-metal sandwichtype structures. In this paper, we present a first-principles calculation, which is based on the methods of Xue-Datta-Ratner \cite{xue1,xue2,xue3}, on the conductance of molecular wires including   benzenedithiol (BDT), benzenedimethanethiol (BDMT), hexanedithiol (HDT), octanedithiol (ODT) and decanedithiol (DDT). An atomically-terminated extended molecule model is used and our calculations are in qualitative agreement with the experimental results \cite{tao1,tao2,tao3}. We also try to analyze the influence from the delocalization of molecular orbitals on the transport properties of molecular junctions. 

\section{Methods}

We use the matrix Green's function method combined with hybrid tight-binding / density functional theory (DFT) for modeling molecular devices, which was firstly introduced by Datta et. al. \cite{dattabook,xue5}. Since what we now consider is the transport properties under equilibrium, the most important thing is to calculate electronic structures of metal-molecule-metal junctions. This problem is the generalization of the familiar chemisorption where a single isolated molecule is considered to be adsorbed onto one metal surface. The molecular device can be divided into two different parts: ``extended molecule'' region (the molecule and the neighbour metallic atoms on the electrodes perturbed by the adsorption of the molecule) and ``contact region''(the unperturbed part of the electrodes) \cite{emberly1,xue5}. This separation provides a convenience to treat the two parts by well-established techniques independently. The solution of the total device system is based on Green's Function method, which supplements the coupling of the molecule and electrodes via the concept of self-energy. The retarded Green's function $G^R$ of the extended molecule is given by
\begin{equation}
	G^R=(ES-F-\Sigma_1^R-\Sigma_2^R)^{-1}
\end{equation} 
	where $E$ is energy, $S$ is the overlap matrix, $F$ is the Fock matrix of the isolated extended molecule. $F$ and $S$ can be obtained through DFT calculation, which is performed using GAUSSIAN 03 program package \cite{g03}. The self-energy matrices $\Sigma_1^R$ and $\Sigma_2^R$  can be written as
\begin{equation}\Sigma_1^R=\tau^{\dagger}_1 g_1^R \tau_1 \ \  and \ \ \Sigma_2^R=\tau_2^{\dagger} g_2^R \tau_2\end{equation}
where $\tau_{1,2}$ is the coupling between the extended molecule and electrodes, and $g_{1,2}$ is  the surface Green's function of the electrodes. 

The computation of the surface Green's function of gold electrodes and the coupling matrices is described using tight-binding scheme with the parameters given by Papaconstantopoulos \cite{papa}. Here we suppose that the gold electrode is semi-infinite Au(111) surface and calculate the Green's function layer by layer with an iteration procedure \cite{xue5}.  Then we can obtain the transmission function through the molecule by the following equation
\begin{equation}T(E)=Tr(\Gamma_1 G^R \Gamma_2 G^A)\end{equation}
	where 
	\begin{equation}\Gamma_{1,2}=i(\Sigma_{1,2}^R-\Sigma_{1,2}^A)\end{equation}
		Since what we concern is the conductance under equilibrium condition, the relation can be given by
		\begin{equation}G=\frac{2e^2}{h}T(E_f)\end{equation}
			The density of state of the extended molecule $D(E)$ can also be calculated by Green's function
			\begin{equation}D(E)=-\frac{1}{\pi}Tr[Im(G^R S)]\end{equation}
\section{Results and discussions}

\subsection{Optimization and calculation models}

	In the following DFT calculations we have used the Becke's three-parameter hybrid functional using the Lee, Yang, and Parr correlation functional \cite{b3,lyp} (B3LYP). The basis set chosen for $C$, $H$, $S$ atoms is the triply split valence with polarization functions, 6-311G** \cite{6-311g1,6-311g2}. When the $Au$ atoms are included in the calculations, two Los Alamos National Laboratory basis sets \cite{lanl1,lanl2} are used with effective core potentials, the LanL2DZ basis set for the geometry optimization and the LanL1MB basis set for the transport calculations. The use of the pseudopotentials with relativistic corrections has been widely demonstrated to be a good compromise with the alternative use of  full-electron procedures, especially when we deal with relatively large numbers of heavy atoms. It requires less computational effort without loss of accuracy \cite{dftbook}.

Under equilibrium condition, we have considered the electronic transport properties in molecular junctions consisting of benzenedithiol (BDT), benzenedimethanethiol (BDMT), hexanedithiol (HDT), octanedithiol (ODT) and decanedithiol (DDT). As the first step in our calculation, we optimize the molecules by a gold trimer-molecule-gold trimer structure, which is shown in Figure \ref{BDMT_opt}. The optimization scheme is similar to our recent work on  calculating 4,4-bipyridine connnected to gold electrodes \cite{Hou}. The Au-S bond lengths of five optimized metal-molecule-metal structures are $2.32\sim 2.33\AA$. All the optimized bond lengths are consistent with other theoretical data \cite{Ratner2}. On the other hand, the calculation of Kru\"ger et. al. also indicates that the gold trimer-molecule structure is stable after leaving from the gold surface \cite{pull}, which supports our optimization scheme.

Assuming the molecular structure remains unchanged during the pulling process by STM tip, we can use the above optimized structures in the following transport calculations. The contact geometry is always a debating problem in the construction of molecular junctions. Some previous work have already done careful discussions on this issue \cite{pull,switch}. Because different adsorption geometries may correspond to dissimilar local minimal energy points \cite{bilic}, we can not determine the accurate contact by only a simple junction model. The work of Emberly's group shows that the conductance of gold-BDT-gold junction changes little when the geometries of contact vary near zero bias \cite{switch}. Their results indicate that an atomically-contact extended molecule model could be one of the reasonable qualitative descriptions for molecular junctions under equilibrium. Thus we model the junctions by inserting a single gold atom between the end sulfur atom and the gold surface at each contact. The inserted gold atom is settled above the center of 3-fold hollow site on Au(111) surface and maintains the Au-S bond length calculated in the above optimized structures. Three nearest-neighbor atoms on the surface are included in the extended molecule, which is showed as an example in Figure \ref{BDMT_Au4}. Within the range of nearest-neighbour coupling, each side of the outmost 3-fold gold atoms in the extended molecule is coupled to the gold electrode through nine gold atoms in the first layer and seven gold atoms in the second layer. All the gold-gold distance is fixed on $2.88\AA$, which is the bond length of body gold. With the above procedure, we construct all the five gold-molecule-gold systems and show them in Figure \ref{models}. For simplicity, all the structures are treated symmetrically.

\subsection{The alignment of the Fermi level}

Another key point for calculating the conductance in molecular junctions is the position of the Fermi level $E_F$. As what Patrone has pointed out \cite{ups2}, UPS experiments can give direct information about $E_F-E_{HOMO}$, which is an important energy reference in the molecular transport calculation. Because we calculate the electrodes and extended molecule separately, we have to face the difficulty that how to unify the zero energy points of two different methods.  Fortunately, introducing the gap of $E_F-E_{HOMO}$ informed by UPS data, we surpass this problem. With energy translation on the Papaconstantopoulpos diagonal Hamiltonian matrix elements \cite{papa}, we can adjust the Fermi level in gold electrodes and locate it in the suitable place within molecular levels obtained by DFT calculations. Thus, the adjustment of the relative energy scale with respect to the experimental data of UPS \cite{ups1,ups3,ups4,ups5} becomes one of the meaningful procedures in our calculations. In Table 1 we show the calculation results of all the five different molecular junctions compared with the experimental values \cite{tao1,tao2,tao3}. To our satisfaction, we  found in our calculation that the conductance measured in STM experiments qualitatively coincide well with the UPS data via the bridge of $E_F-E_{HOMO}$.  

As what we have shown in Table 1, after considering the UPS data of $E_F-E_{HOMO}$  $0.6eV$ \cite{ups4}, the calculation of BDT's transmission near the Fermi level ($T(E_f)=0.0086$) agrees well with the experimental result $0.011$.  UPS experiments also show that the position of HOMO is larger than $5eV$ from the Fermi level in octandecanethiol monolayers on Au \cite{ups1,ups5}. For the  similar electronic structures of three saturated alkanedithiol molecules (HDT, ODT, DDT), here we assume $5.5eV$ as an median value for the calculation of all the three molecules. The results of $0.85\times 10^{-3}(HDT)$, $3.17\times 10^{-4}(ODT)$, $2.62\times 10^{-5}(DDT)$ could be qualitatively in accordance with the experimental results $1.2\times 10^{-3}(HDT)$, $2.5\times 10^{-4}(ODT)$, $2\times 10^{-5}(DDT)$ \cite{tao1,tao2,tao3}. Although we do not find direct UPS experimental results about BDMT, the measurement of 4,4-(ethynylphenyl)-1-benzenethiol \cite{ups3}, which has a comparable electronic structure with the incorporation of phenyl and carbon-chain group like BDMT, can give us useful enlightenments. Unlike the small value $(0.5\sim 0.6eV)$ in pure $\pi$-bonded molecules \cite{ups2,ups4} and the large value $(>5eV)$ in saturated carbon-chain molecules \cite{ups1,ups5}, the above mixed  electronic structure has a $1.9eV$ gap of $E_F-E_{HOMO}$ \cite{ups3}. It provides the prompt on estimating the $E_F-E_{HOMO}$ value of BDMT. Here we use $2.1eV$ and find that the calculation result $6.79\times 10^{-4}$ is in good agreement with the experimental data $6\times 10^{-4}$. In addition, we consider the influence on the calculation results due to the  uncertainty in UPS experiments. In Table 1, we change $E_F-E_{HOMO}$ informed by UPS data with a value of $\pm 0.2eV$. Our calculation shows that the results are not affected qualitatively. It indicates that introducing UPS experimental information might provide a relatively reliable energy reference for qualitative calculations on molecular conductance under equilibrium.

\begin{center}
\begin{tabular}{ccccc}
	\multicolumn{4}{l}{Table. 1 The used Energy gap of $E_F-E_{HOMO}$ informed by UPS data \cite{ups1,ups3,ups4,ups5} and }\\
	\multicolumn{4}{l}{ the comparison of calculated molecular conductances with the experimental results \cite{tao1,tao2,tao3}}\\ 
	\hline\hline
	&\multicolumn{1}{c}{Energy gap($eV$)}&\multicolumn{2}{c}{Conductance($G_0$)}\\
	\cline{2-4}\\
MOLECULAR WIRE &$E_F-E_{HOMO}$ &CALCULATION & EXPERIMENT\\
\hline
&0.4&$1.14\times 10^{-2}$ &\\
BDT&0.6&$0.86\times 10^{-2}$ &$1.1\times 10^{-2}$\\
&0.8&$0.77\times 10^{-2}$ &\\
\hline
&1.9&$9.69\times 10^{-4}$ &\\
BDMT&2.1&$6.79\times 10^{-4}$ &$6\times 10^{-4}$\\
&2.3& $5.24\times 10^{-4}$ &\\
\hline
&5.3&$0.60\times 10^{-3}$ &\\
HDT&5.5&$0.85\times 10^{-3}$ &$1.2\times 10^{-3}$\\
&5.7& $1.3\times 10^{-3}$ &\\
\hline
&5.3&$2.29\times 10^{-4}$ &\\
ODT&5.5&$3.17\times 10^{-4}$ &$2.5\times 10^{-4}$\\
&5.7& $4.65\times 10^{-4}$ &\\
\hline
&5.3&$2.27\times 10^{-5}$ &\\
DDT&5.5&$2.62\times 10^{-5}$ &$2\times 10^{-5}$\\
&5.7& $3.21\times 10^{-5}$ &\\
\hline\hline
\end{tabular}
\end{center}

\subsection{Analysis on density of states (DOS)}
The above results indicate the particular molecular electronic structures would associate with the alignment of the Fermi level in the gap between HOMO and LUMO. Basically, the energy position of the Fermi level and the degree of delocalization in molecular junctions are considered as two major factors which impact the transport properties contributed from  molecular orbitals \cite{PRL88256803}. In order to make a better comprehension on such an issue, we do some analysis on DOS of molecular junctions. For the consideration of charge transfer between gold electrodes and the molecule, here we use a central cluster (Au-molecule-Au): one gold atom on each side of the molecule is included in our calculation. Mulliken's population analysis shows only a small charge transfer ($0.12\sim 0.17e$) between the central cluster and the remained gold electrode parts in the above gold-molecule-gold models. So we believe that the main charge transfer has been properly described with a single gold atom connected to each side of the molecule. 

In Figure \ref{BDT_cal}, we show the transmission under zero bias and the DOS of different parts in Au-BDT-Au as a function of energy. We find that the contribution from Au is significant in the gap between molecular levels. In the junction of BDT, about $75\%$ of DOS near the Fermi level is from gold. And in the other four junctions, the percentage of gold's contribution  is around $90\%$. It means that gold electrodes might play an important role in molecular transport. When the coupling between molecules and electrodes is weak (in gold-HDT-gold, gold-ODT-gold, gold-DDT-gold junctions), the DOS of gold is mostly localized around the gold-molecule interface and leads to a tunneling behavior. Cui et.al. have discussed thoroughly on this subject of the transport in alkanedithiol molecular junctions by a tunneling model \cite{cui}. Their results are in accordance with our calculation shown in Figure \ref{XDT}. In addition, Kaun's calculation suggests that transport accurs through HOMO \cite{kaun} while our results show transport might accur through LUMO in alkanedithiol-gold junctions.  On the other hand, the conductance in gold-BDT-gold junction is larger than the one in alkanedithiol junctions. The existence of $\pi$ bond results in a strong hybridization of molecular states and gold's state, which implies a more complex transmission mechanism far beyond tunneling. 

We can also find additional information about transmission in the projected DOS (PDOS) on different molecular orbitals of Au-molecule-Au clusters. Here we reference the scheme of Heurich et. al.\cite{PRL88256803}. We choose the clusters including Au atoms for the decomposition on molecular orbitals. Such a scheme can directly reflect on the change of molecular orbitals induced by the adsorption on electrodes.  We define new molecular orbitals after the hybridization of molecular states and gold' state in Au-molecule-Au clusters, which are meaningful but different from the ones in isolated molecules. Figure \ref{BDT_a}, Figure \ref{BDMT_a}, and Figure \ref{HDT_a} respectively show the PDOS of several representative molecular orbitals in Au-BDT-Au, Au-BDMT-Au and Au-HDT-Au.  The corresponding orbital-shape plots are shown in Figure \ref{BDT_b}, Figure \ref{BDMT_b} and Figure \ref{HDT_b}. According to these figures, the contribution of a special molecular orbital (MO) to DOS strongly depends on its charge-density distribution on this MO. The extended and well-coupled MOs, such as the HOMO, LUMO and LUMO+$1$ in Au-BDT-Au, have a significant weight in DOS. After coupled with gold through 6s atomic orbital, these molecular orbitals extended widely ($>6eV$). The HOMO and LUMO+1 are even partly hybridized. The delocalization of $\pi$-bonded electronic structure is the main reason for the existence of such MOs. On the contrary, the localized PDOS always derives from the confined charge-density on this MO. The orbital shape of Au-HDT-Au in Figure \ref{HDT_b} shows a distinct feature of localization about $\sigma$-bonded molecules. Figure \ref{BDMT_level} also shows an obvious influence of -(CH$_2$)- structure on transmission. It isolates the coupling between electrodes  and the $\pi$ bond in BDMT and results in a localized DOS distribution of most MOs. 

\section{Conclusion}
In this work, the conductances of a series of gold-dithiol molecule-gold junctions are calculated by matrix Green's  function method combined with hybrid tight-binding/ density functional theory. The atomically-contacted extended molecule model is used for the calculation. After determining  the position of the Fermi level by the energy reference according to the results from UPS experiments, the final calculation of molecular conductances qualitatively agree with the experimental values obtained by Tao and co-workers \cite{tao1,tao2,tao3}. In our calculation, the influence by the uncertainty of UPS data is also considered. Our results show that the first-principles method based on Green's function method can qualitatively predict the conductance properties near low bias under the instruction of UPS experiments.  

{\ }

{\noindent\bf\Large Acknowledgments}
	\\
	\\
The authors thank Huaiyu Wang for helpful comments. This project was supported by the National Natural Science Foundation  of China (NO. 90207009, 90206048, 90406014).

\begin{fig}
\centering
\includegraphics[width=6.5cm]{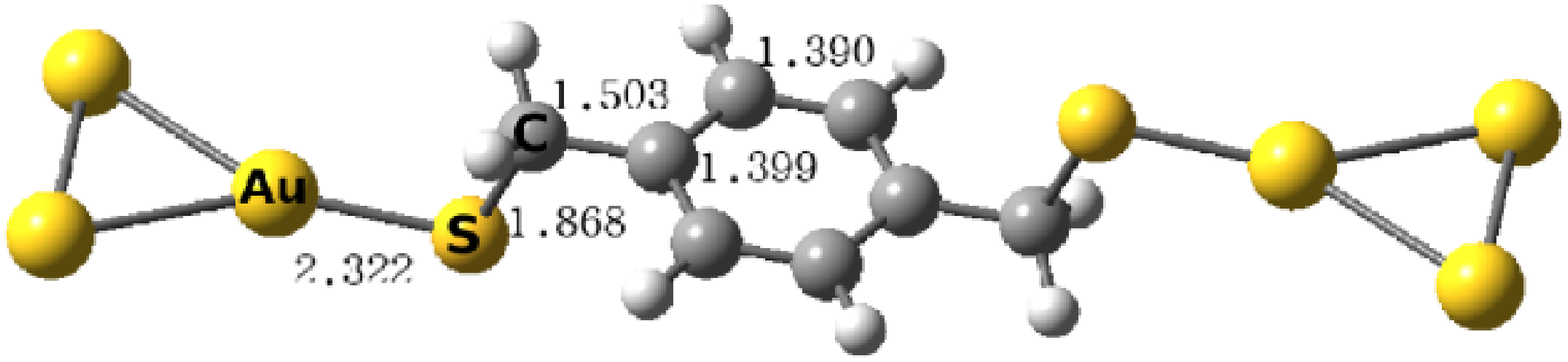}
\caption{\footnotesize{ The optimized geometry for Au3-BDMT-Au3. The optimization method is B3LYP/6-311G** + LANL2DZ.} }
\label{BDMT_opt}
\end{fig}
\begin{fig}
\centering
\includegraphics[width=6.5cm]{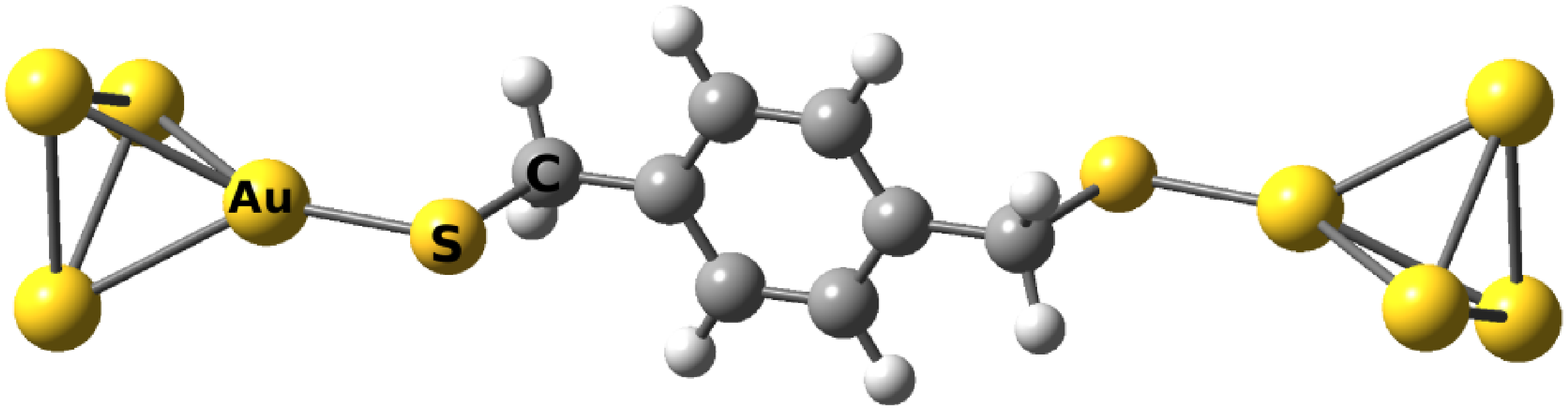}
\caption{\footnotesize{ The example of extended molecule: Au4-BDMT-Au4} }
\label{BDMT_Au4}
\end{fig}
\begin{fig}
\centering
\includegraphics[width=6.5cm]{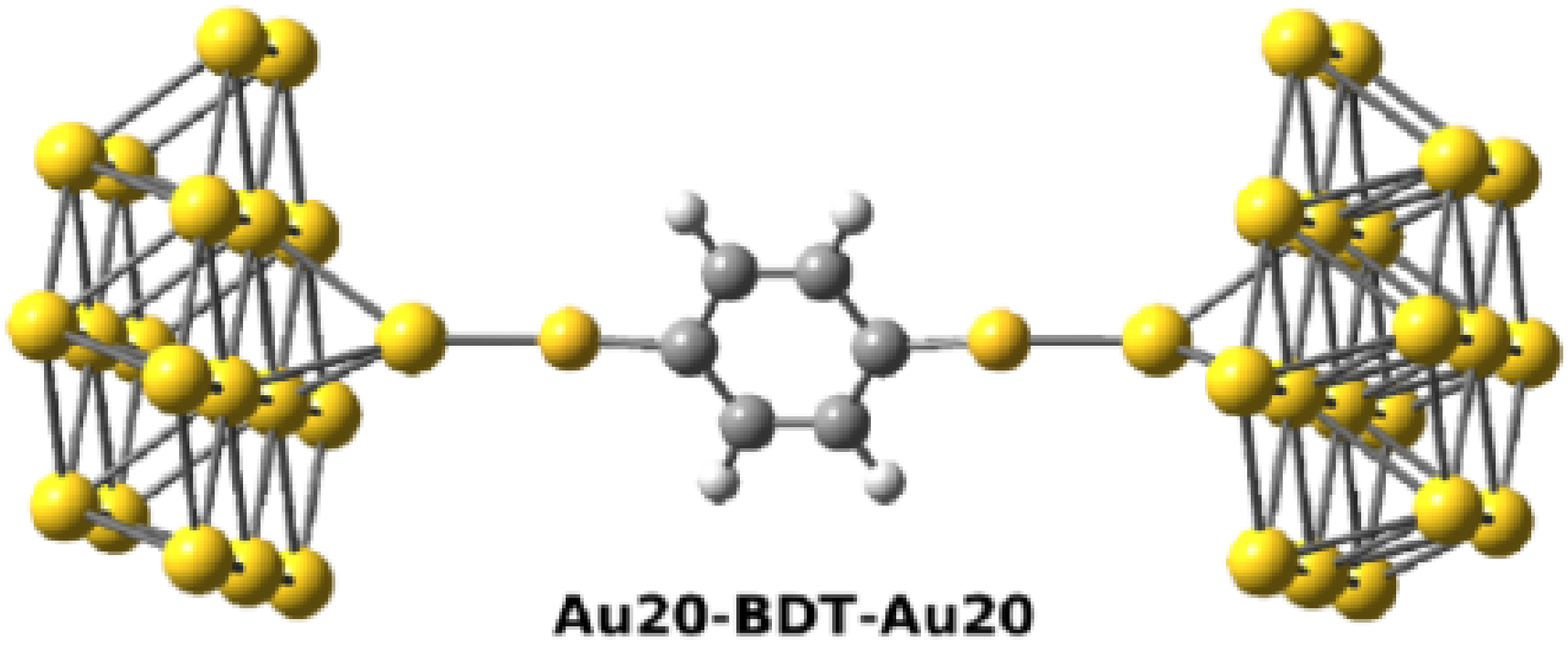}
\includegraphics[width=6.5cm]{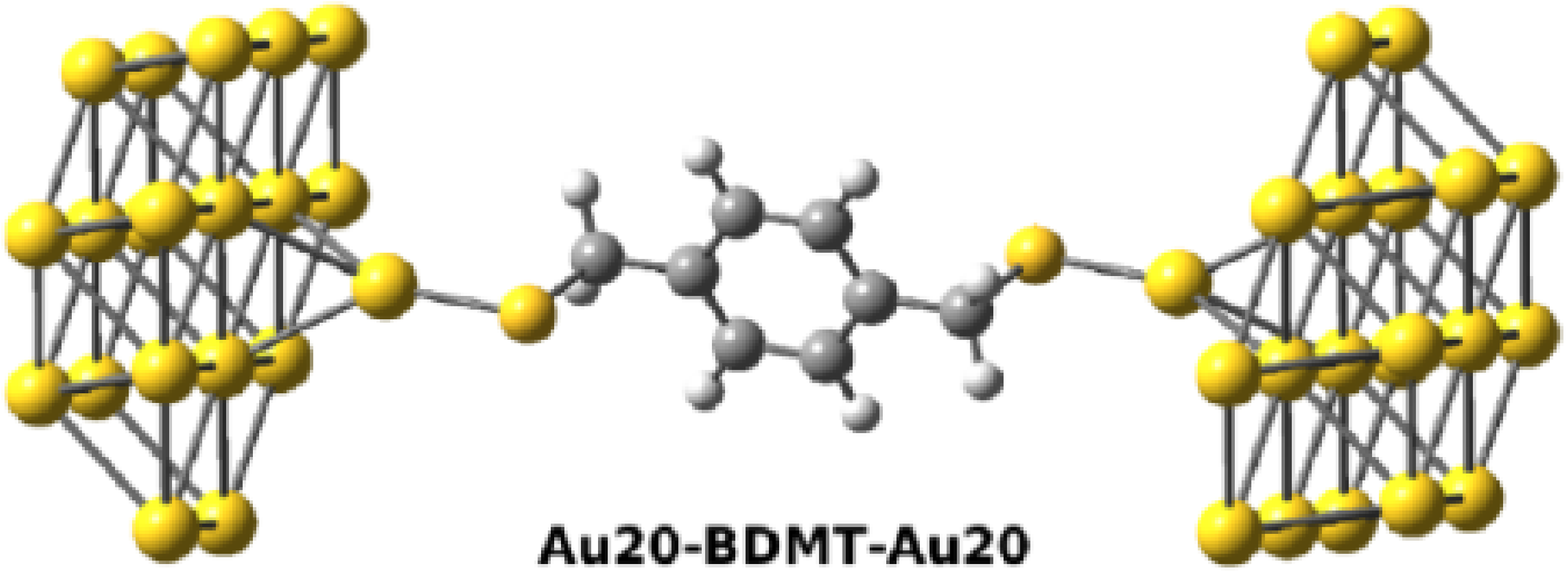}
\includegraphics[width=6.5cm]{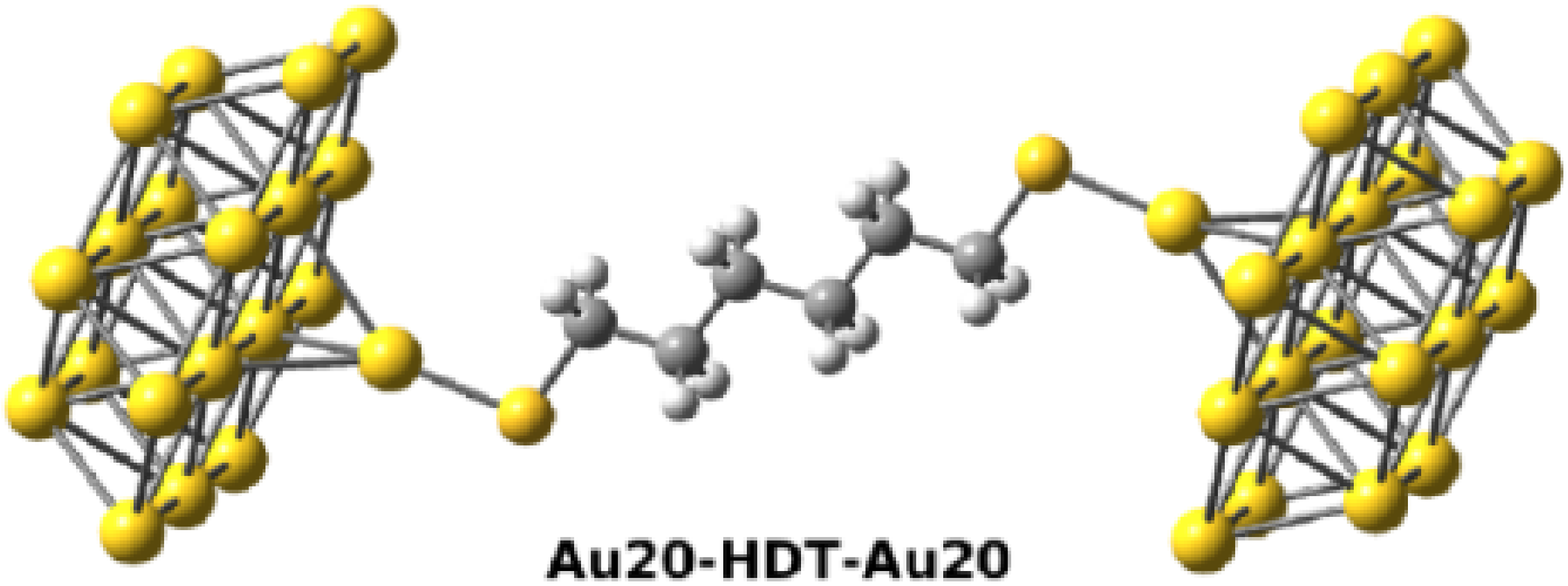}
\includegraphics[width=6.5cm]{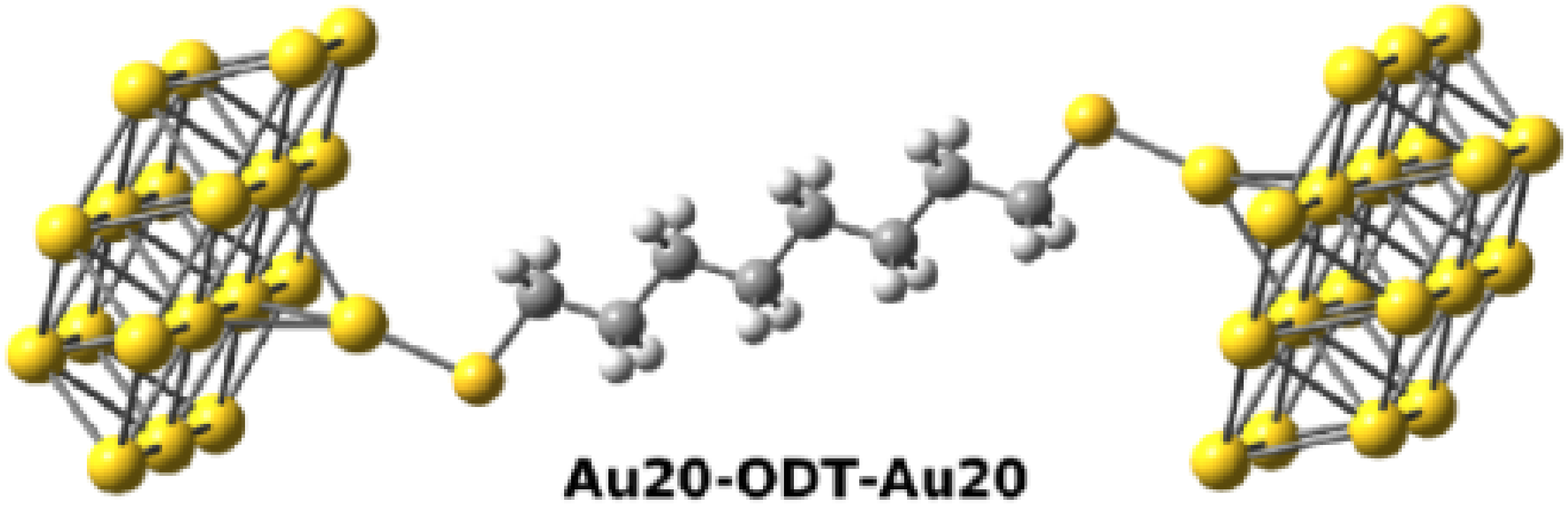}
\includegraphics[width=6.5cm]{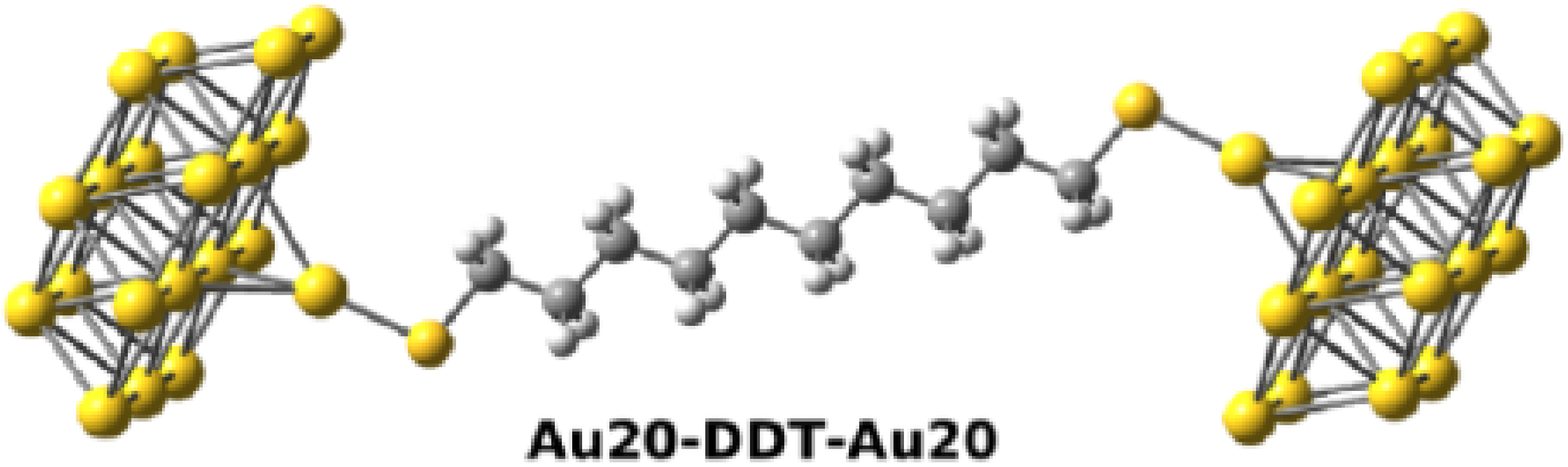}
\caption{\footnotesize{ Au20-molecule-Au20 clusters used in the calculations of conductances with B3LYP/6-311G** + LANL1MB. The molecules are benzenedithiol (BDT), benzenedimethanethiol (BDMT), hexanedithiol (HDT), octanedithiol (ODT) and decanedithiol (DDT).} }
\label{models}
\end{fig}

\begin{fig}
\centering
\includegraphics[width=6.5cm]{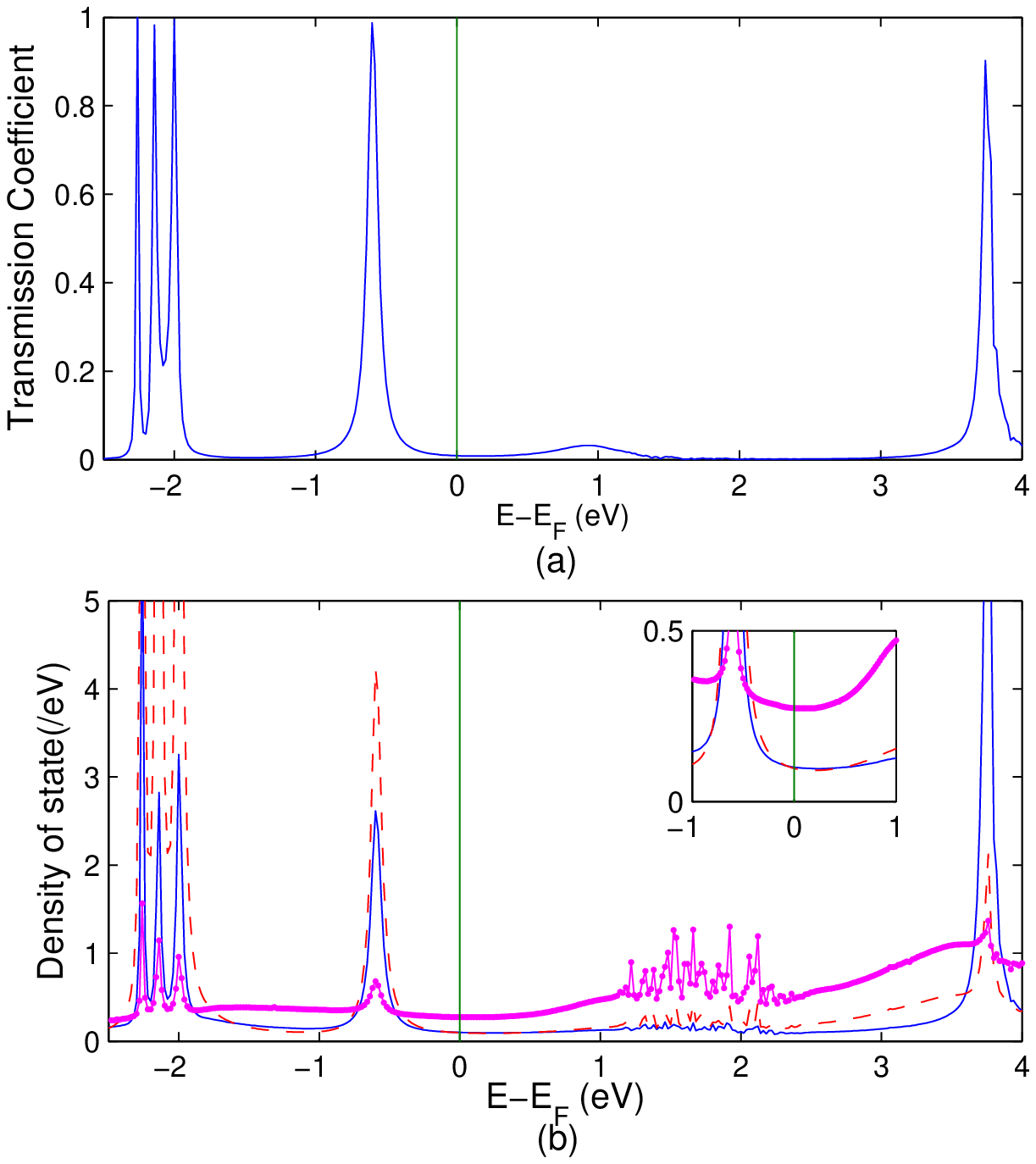}
\caption{\footnotesize{ (a)The transmission coefficient (b) The contribution to total density of state (TDOS) of Au-BDT-Au from different parts. The blue solid line is the part from phenyl ring. The red dashed line is the part from sulfur. The magenta thick line is the part from gold. In the inset of the amplified figure of DOS, we can see the DOS near the Fermi level. } }
\label{BDT_cal}
\end{fig}

\begin{fig}
\subfigure[]{
\begin{minipage}[b]{0.4\textwidth}
\centering\label{BDMT_transmission}
\includegraphics[width=6.5cm]{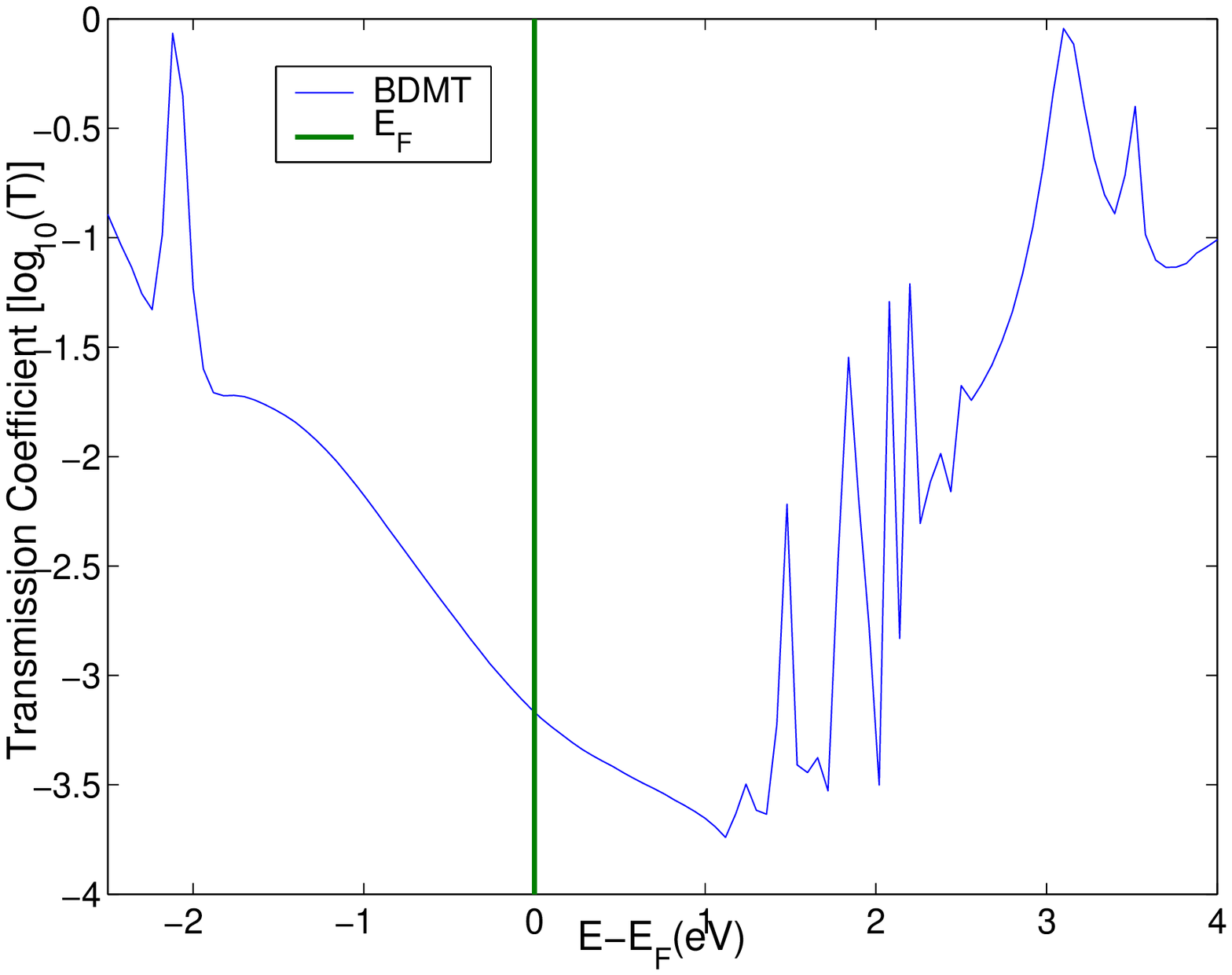}
\label{BDT_transmission}
\end{minipage}}%
\subfigure[]{
\begin{minipage}[b]{0.6\textwidth}
\centering\label{alkanethiol}
\includegraphics[width=6.5cm]{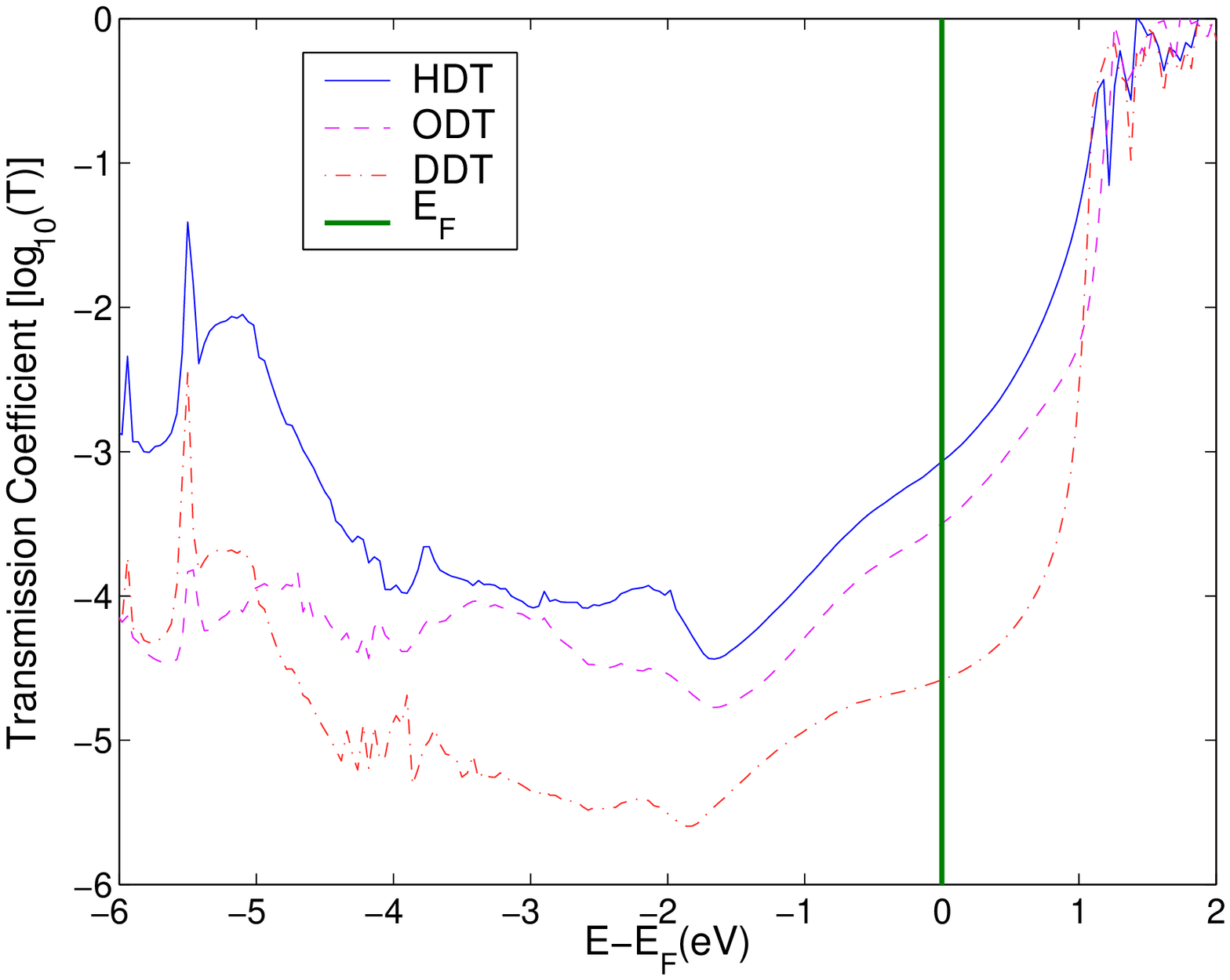}
\label{XDT}
\end{minipage}}%
\caption{\footnotesize{ (a) Transmission coefficient of Au-BDMT-Au junction. (b) Transmission coefficient of Au-HDT-Au, Au-ODT-Au, Au-DDT-Au junctions.}}
\end{fig}

\begin{fig}
\subfigure[]{
\begin{minipage}[t]{0.4\textwidth}
\centering\label{BDT_a}
\includegraphics[width=6.5cm]{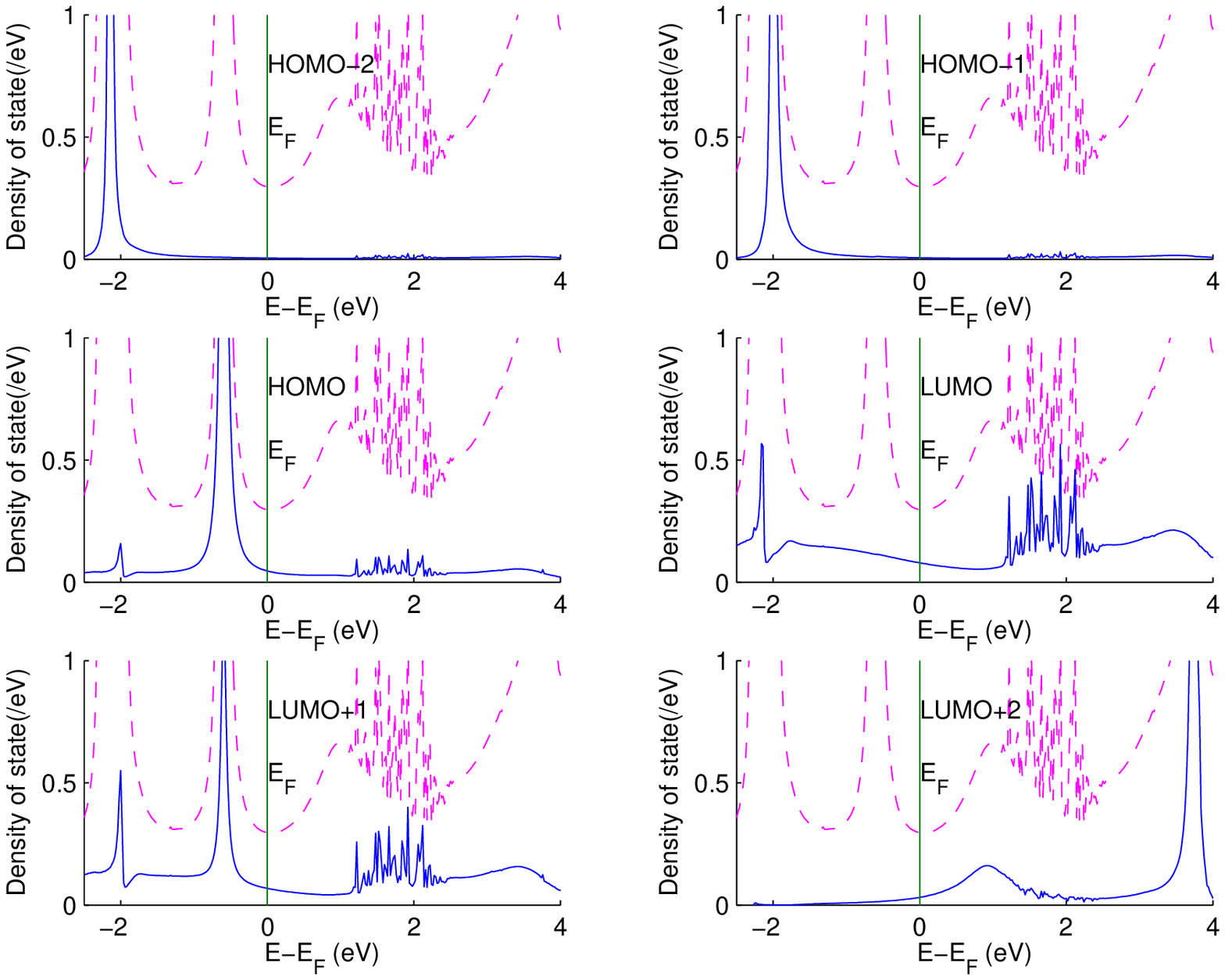}
\end{minipage}}%
\subfigure[]{
\begin{minipage}[t]{0.6\textwidth}
\centering\label{BDT_b}
\includegraphics[width=6.5cm]{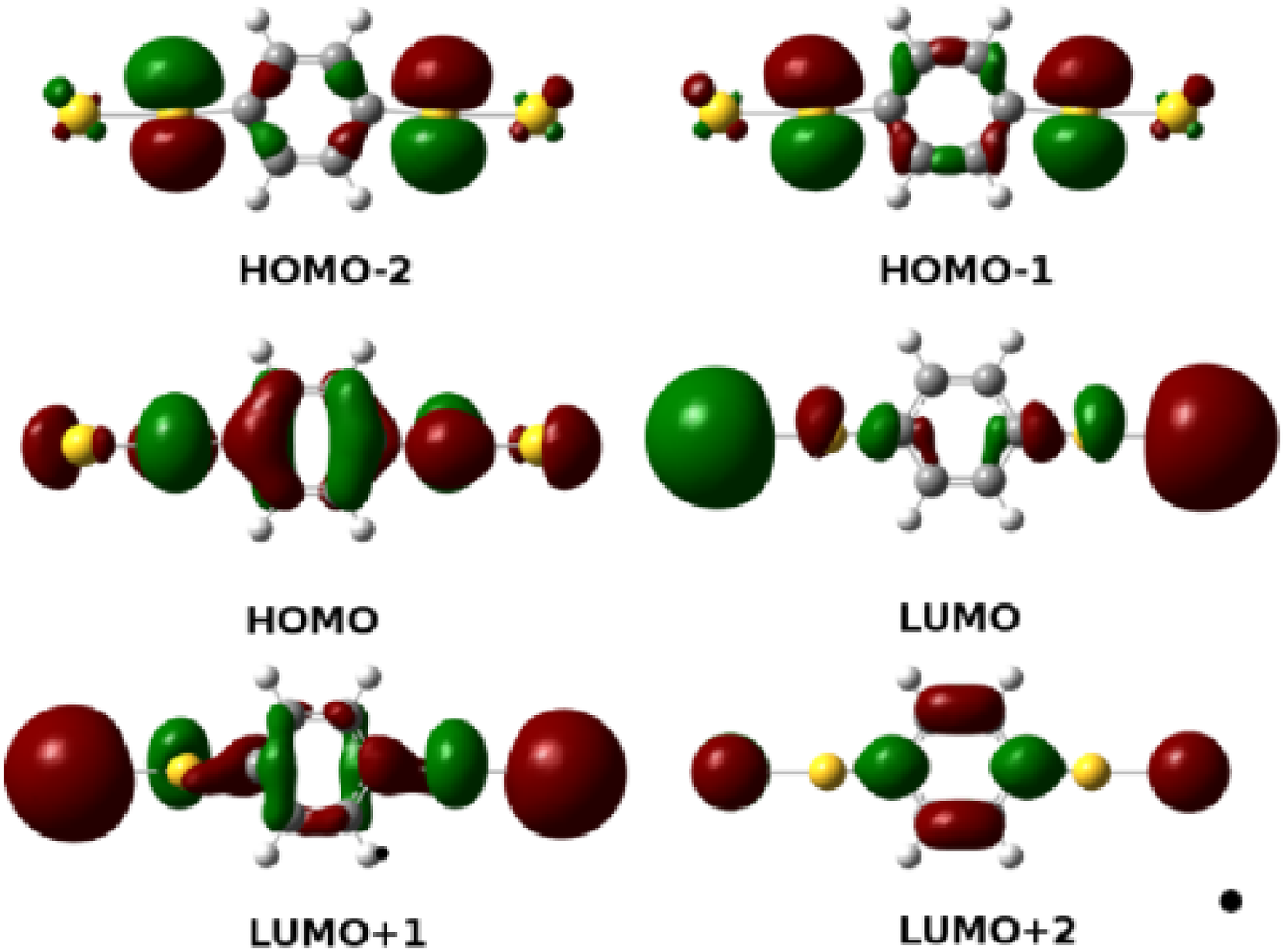}
\end{minipage}}%

\caption{\footnotesize{ (a) The projected density of states (solid line) of six molecular orbitals of Au-BDT-Au. The dashed line is the total density of state of Au-BDT-Au. (b) The orbital shapes of the corresponding molecular orbitals in  (a)} }
\label{BDT_level}
\end{fig}
\begin{fig}
\subfigure[]{
\begin{minipage}[b]{0.4\textwidth}
\centering\label{BDMT_a}
\includegraphics[width=6.5cm]{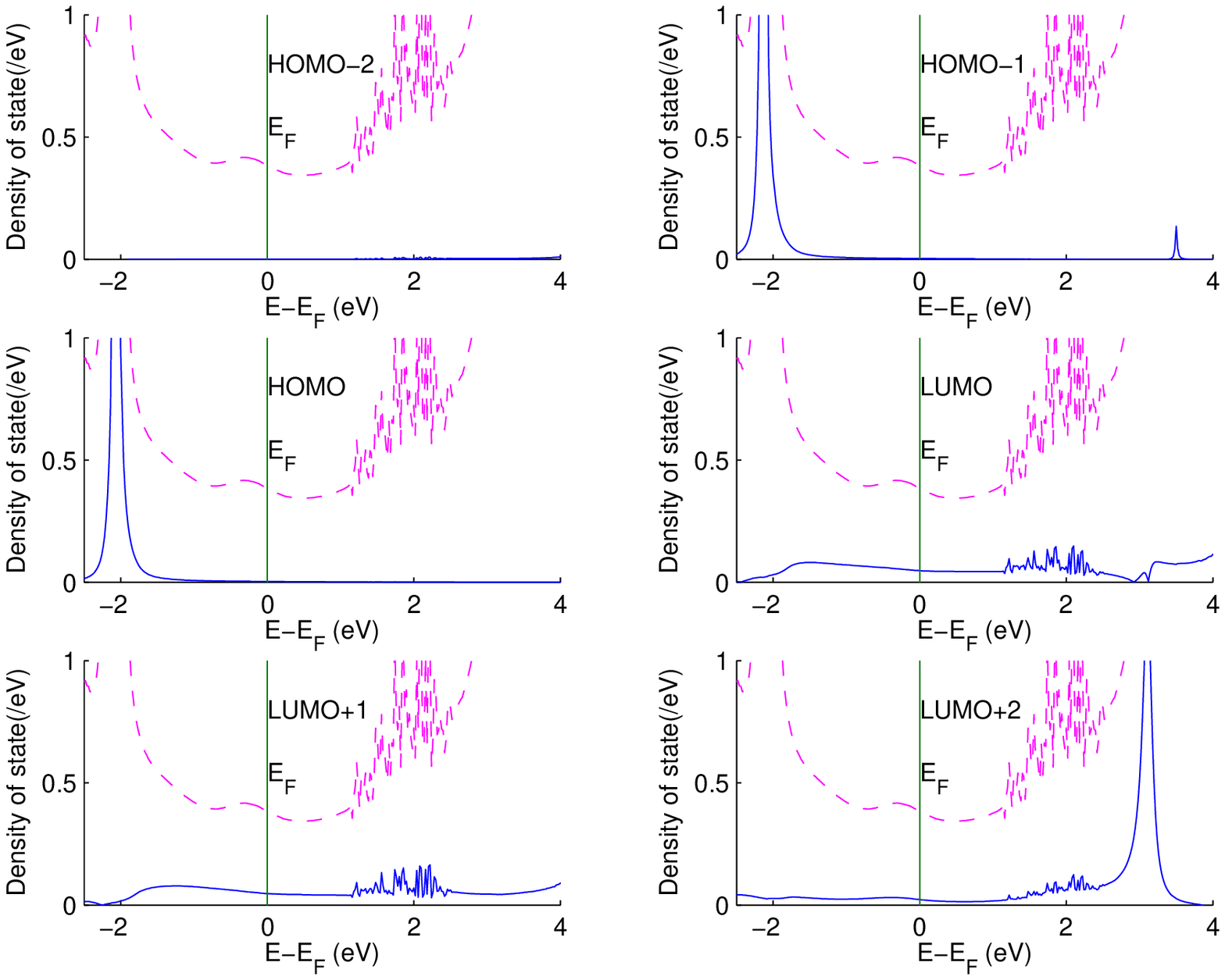}
\end{minipage}}%
\subfigure[]{
\begin{minipage}[b]{0.6\textwidth}
\centering\label{BDMT_b}
\includegraphics[width=6.5cm]{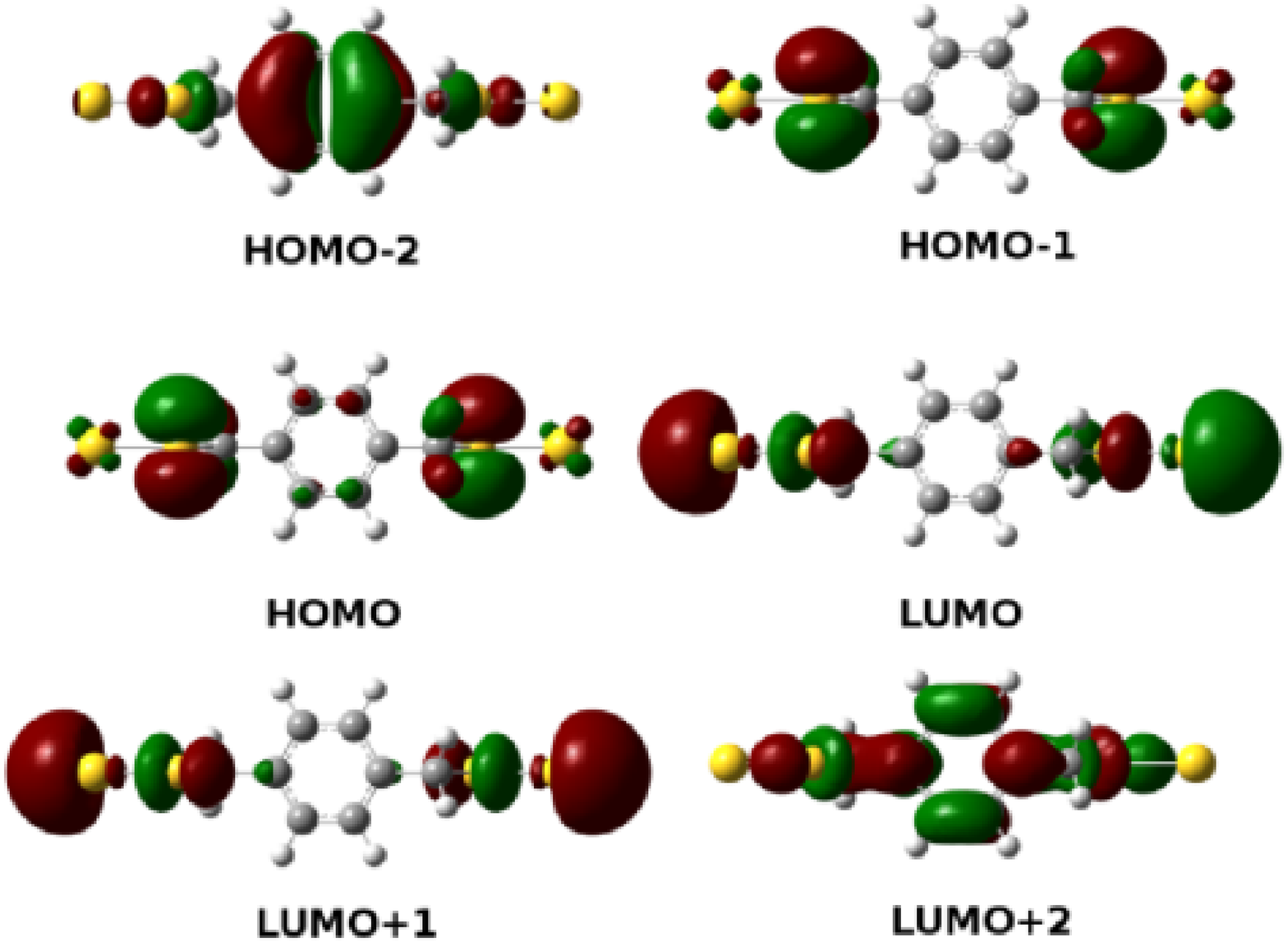}
\end{minipage}}%
\caption{\footnotesize{ (a) The projected density of states (solid line) of six  molecular orbitals of Au-BDMT-Au. The dashed line is the total density of state of Au-BDMT-Au. (b) The orbital shapes of the corresponding molecular orbitals in (a)} }
\label{BDMT_level}
\end{fig}
\begin{fig}
\subfigure[]{
\begin{minipage}[b]{0.4\textwidth}
\centering\label{HDT_a}
\includegraphics[width=6.5cm]{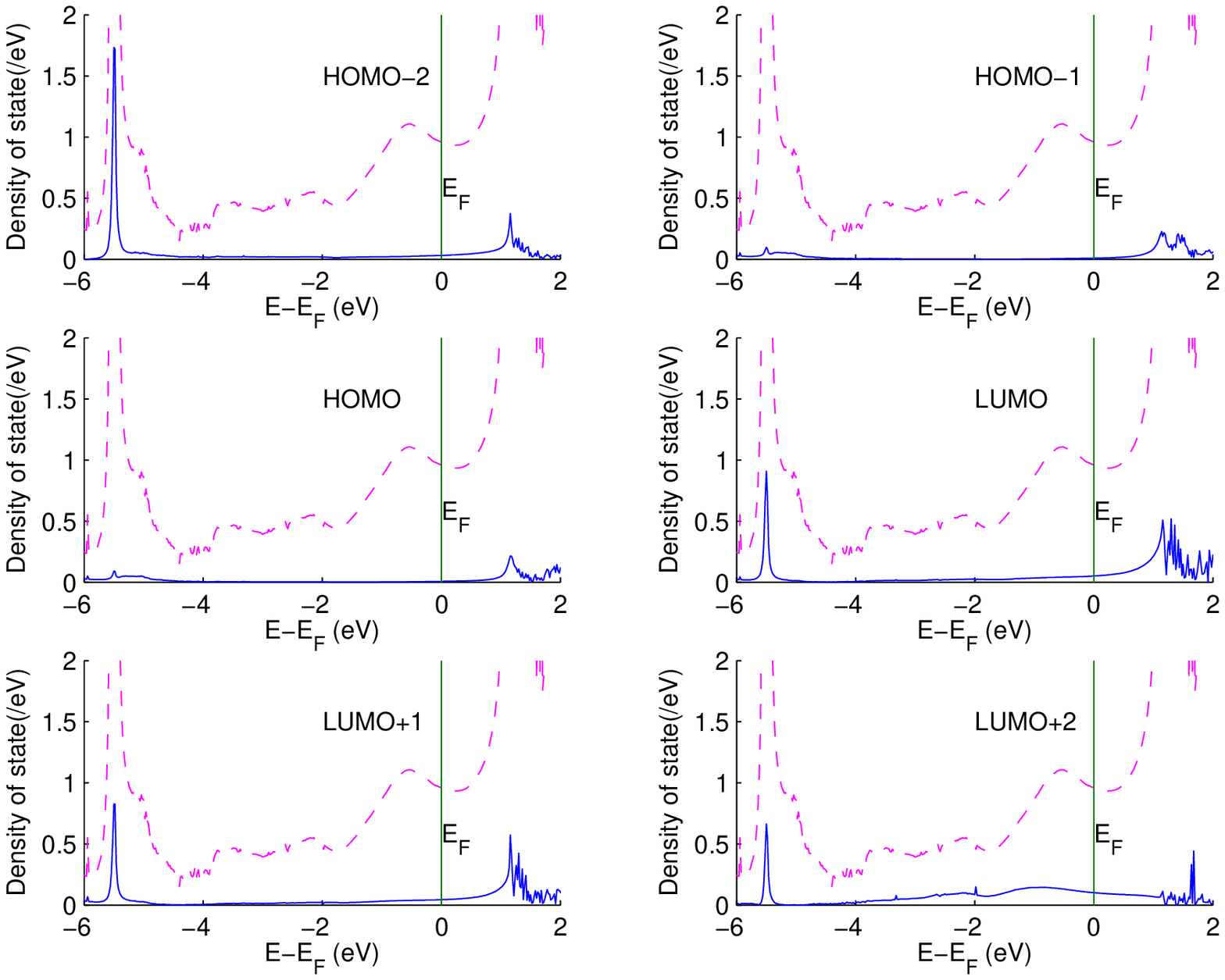}
\end{minipage}}%
\subfigure[]{
\begin{minipage}[b]{0.6\textwidth}
\centering\label{HDT_b}
\includegraphics[width=6.5cm]{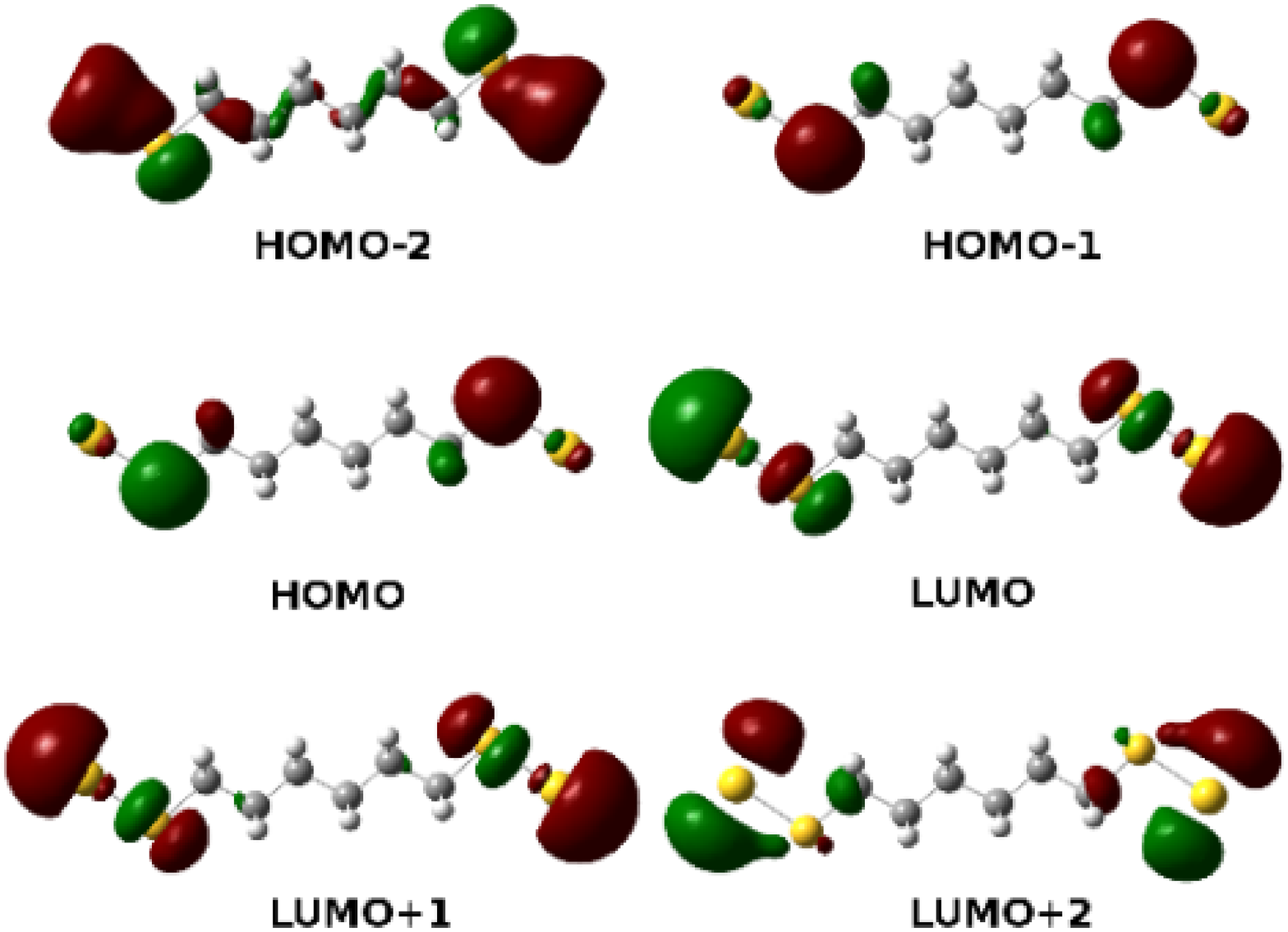}
\end{minipage}}%
\caption{\footnotesize{ (a) The projected density of states (solid line) of six molecular orbitals of Au-HDT-Au. The dashed line is the total density of state of Au-HDT-Au. (b) The orbital shapes of the corresponding molecular orbitals in (a)} }
\label{HDT_level}
\end{fig}

\small{

}
\end{document}